\documentclass[
    reprint,
    aps,
    amsmath,
    amssymb,
    superscriptaddress]{revtex4-2}
\usepackage{graphicx} 
\usepackage{dcolumn}  
\usepackage{bm}       
\usepackage{amsfonts}
\usepackage{dsfont}
\usepackage{amsmath}
\usepackage{url}

\usepackage{ulem}
\usepackage{color}

\begin{document}

\title{Valley polarization control in WSe$_2$ monolayer by a single-cycle laser pulse}

\author{Arqum Hashmi}
\affiliation{%
Kansai Photon Science Institute, National Institutes for Quantum  Science and Technology (QST), Kyoto 619-0215, Japan }
            
\author{Shunsuke Yamada}
\affiliation{%
Center for Computational Sciences, University of Tsukuba, Tsukuba 305-8577, Japan}

\author{Atsushi Yamada}
\affiliation{%
Center for Computational Sciences, University of Tsukuba, Tsukuba 305-8577, Japan}

\author{Kazuhiro Yabana}
\affiliation{%
Center for Computational Sciences, University of Tsukuba, Tsukuba 305-8577, Japan}

\author{Tomohito Otobe}
\email[]{otobe.tomohito@qst.go.jp}
\affiliation{%
Kansai Photon Science Institute, National Institutes for Quantum  Science and Technology (QST), Kyoto 619-0215, Japan }

\date{\today}

\begin{abstract}
The valley degree of freedom in two-dimensional materials provides an opportunity to extend the functionalities of valleytronics devices. Very short valley lifetimes demand the ultrafast control of valley pseudospin. Here, we theoretically demonstrate the control of valley pseudospin in WSe$_{2}$ monolayer by single-cycle linearly polarized laser pulse. We use the asymmetric electric field controlled by the carrier-envelope phase (CEP) to make the valley polarization between $K$ and $K^{'}$-point in the Brillouin zone (BZ). Time-dependent density functional theory with spin-orbit interaction reveals that no valley asymmetry and its CEP dependence is observed within the linear-optical limit. In the nonlinear-optical regime, linearly polarized pulse induces a high degree of valley polarization and this polarization is robust against the field strength. Valley polarization strongly depends and oscillates as a function of CEP. The carrier density distribution forms nodes as the laser intensity increases, our results indicate that the position of the carrier density in the BZ can be controlled by the laser intensity. From the analysis by the massive Dirac Hamiltonian model, the nodes of the carrier density can be attributed to the Landau-Zener-St\"uckelberg interference of wave packets of the electron wave function. 
\end{abstract}

\maketitle


\section{\label{sec:level1}INTRODUCTION \protect }

 Mechanical exfoliation of atomically thin layers by scotch tape from van der Waals bulk crystals has opened up new opportunities for the design of nanoscale quantum materials~\cite{osada2012two, chhowalla2013chemistry}. Specifically, the realization of monolayer graphene in 2004~\cite{novoselov2004electric} has ignited extensive research efforts on two-dimensional (2D) layered materials. 2D materials exhibit unique mechanical, optical, and electronic properties compared to their bulk counterparts~\cite{kang20202d}. Owing to their extraordinary physical properties, the study of 2D monolayers has now been established as an emerging field. For instance, one can find numerous reports on graphene~\cite{novoselov2012roadmap}, silicene~\cite{tao2015silicene}, transition metal dichalcogenides (TMDC)~\cite{wang2012electronics}, and phosphorene~\cite{xia2014rediscovering}. 2D materials have a wide range of applications in the field of electronics and an evolving field of optoelectronics. They feature strong light-matter interaction~~\cite{kang20202d}, ultrafast broadband optical response~\cite{tan20202d}, and large optical nonlinearity~\cite{wen2019nonlinear}, thus have a great potential for optoelectronic applications such as photodetectors, tunneling and imaging devices~\cite{chhowalla2013chemistry, kang20202d, zhou20182d}. 

2D materials are classified as magnetic and nonmagnetic semiconductors, topological insulators, metals, and half metals. Within a wide class of 2D materials family, materials with broken inversion symmetry, such as TMDC monolayers are getting special attention~\cite{schaibley2016valleytronics}. The lack of inversion symmetry in TMDC monolayer induces a novel Zeeman type spin splitting which results in two degenerate yet inequivalent valleys in the band structure~\cite{yao2008valley, zhu2011giant, xiao2012coupled, yuan2013zeeman}. Valleys are local minima that correspond to different crystal momentum in the reciprocal space. Spin-orbit coupling (SOC) lifts the spin degeneracy in both valleys and the opposite spin angular momenta appear in two valleys owing to the time-reversal symmetry. Thus, the spin-valley locking and the interplay of two inequivalent valleys give rise to valley-dependent optical selection rules~\cite{xu2014spin}. Manipulation of valley pseudospin thus becomes a central theme in the field of valleytronics. Several methods have been proposed to achieve transient valley polarization such as optical excitations~\cite{zeng2012valley, mak2012control, yuan2014generation}, by applying an external magnetic field and magnetic proximity effect induced by the substrate~\cite{sanchez2016valley, ye2016electrical, norden2019giant}. However, due to the several practical limitations and very short valley lifetimes 10${}^{3}$-10${}^{6}$ femtoseconds (fs)~\cite{vitale2018valleytronics}, ultrafast control of valley selection on fs time scales is in urgent need.

Recent experimental demonstration of intense terahertz pulses driven sub-cycle control of valley dynamics has opened a way to manipulate the valley pseudospin that is switchable within few fs~\cite{langer2018lightwave}. Recently, A. J-Galan et al. have also shown to control the valley excitation by using non-resonant driving fields on a fs timescale~\cite{jimenez2020lightwave}. Moreover, valley polarization using few-cycle linearly polarized pulses with the controlled carrier-envelope phase (CEP) has also been proposed by using the density matrix approach~\cite{jimenez2021sub}. 

In this work, we would like to investigate the linear polarized single-cycle laser pulse control of valley pseudospin in WSe$_{2}$ monolayer employing time-dependent density functional theory (TDDFT). TDDFT can describe electron dynamics under intense laser field without any empirical parameters~\cite{otobe2008first}.  We have developed the program for the electron and electro-magnetic field dynamics, open-source package Scalable Ab-initio Light-Matter simulator for Optics and Nanoscience (SALMON)~\cite{noda2019salmon,Salmon:Online}, which employs the TDDFT.  
We have implemented the SOC with non-collinear local spin density~\cite{von1972local,oda1998fully,theurich2001self} to the SALMON to describe the spin-dependent electron dynamics.

Laser intensity and carrier-envelope phase (CEP) dependence of the valley polarization is studied. Linearly polarized single-cycle pulse induces a high degree of valley polarization. Moreover, the valley polarization is robust against the field strength although it oscillates as a function of CEP. In the strong field regime, we found that the distinct node formation  of carrier density in the Bloch phase space induced by the quantum interference that can be tuned by the laser intensity. Our results demonstrate the single optical cycle control of valley pseudospin by linear polarized laser pulses.

\section{THEORETICAL FORMALISM}

\subsection{TDDFT}

We use the 2D approximation method that describes the electron dynamics and light propagation in extremely thin layers at normal incidence~\cite{yamada2018time, yamada2021determining}. Here, we briefly describe the theoretical formalism for this method. 

The polarization and propagation directions for light pulses are taken along the $x$ axis and $z$ axis, respectively. 
Also, we assume that the thin layer is in the $xy$ plane.
We consider only the $x$ component of vector fields and omit the label ``$x$".
By using the Maxwell equations, we can describe the propagation of macroscopic electromagnetic fields in the form of the vector potential $A(z,t)$ as,
\begin{equation} \label{GrindEQ__2_} 
\frac{1}{c^2}\ \frac{{\partial }^2A(z,t)}{{\partial t}^2}-\ \frac{{\partial }^2A\left(z,t\right)}{{\partial z}^2}=\ \frac{4\pi }{c}J(z,t),
\end{equation} 
where $J(z,t)$ is the macroscopic current density in a thin layer.
For an atomic monolayer material, the macroscopic electric current density in Eq. (1) can be expressed as  
\begin{equation} \label{GrindEQ__3_} 
J\left(z,t\right)\approx  \delta( z)  J_{\rm 2D}(t),
\end{equation} 
where $J_{\rm 2D}(t)$ is 2D current density  of the monolayer. 
We deal with it as a boundary value problem where reflected (transmitted) fields can be determined by the connection conditions at z = 0. 
From Eq.~(\ref{GrindEQ__3_}), we obtain the continuity equation of $A(z,t)$ at $z=0$ as follows
\begin{equation}
A(z=0,t)=A^{\rm (t)}(t) =  A^{\rm (i)}(t) + A^{\rm (r)}(t),
\end{equation}
where the $A^{\rm (i)}$,  $A^{\rm (r)}$, and  $A^{\rm (t)}$ are the incident, reflected, and transmitted fields, respectively.
From the Maxwell equation~(\ref{GrindEQ__2_}) and Eq.~(\ref{GrindEQ__3_}), we get the basic equation of the 2D approximation method,
\begin{equation} \label{GrindEQ__5_} 
\frac{dA^{\rm (t)} }{dt}  = \frac{dA^{\rm (i)}}{dt}+ \ 2\pi  J_{\rm 2D}\left[A^{\rm (t)}\right](t).
\end{equation} 
Here, $J_{\rm 2D}\left[A^{\rm (t)}\right](t) $ is the 2D current density that is determined by the vector potential at $z=0$ and it is equal to $A^{\rm (t)}(t)$. 
By using the velocity gauge~\cite{bertsch2000real}, the time-dependent Kohn-Sham (TDKS) equation using Bloch orbitals $u_{b,{\bf k}}({\bf r},t)$ ($b$ is the band index and ${\bf k}$ is the 2D crystal momentum of the thin layer) is described as 
\begin{equation} \label{GrindEQ__4_} 
\begin{split}
i\hbar  \frac{\partial }{\partial t}u_{b, {\bf k}}({\bf r},t)=\Big[\frac{1}{2m}{\left(-i\hbar \nabla +\hbar{\bf k} + \frac{e}{c} \hat{\bf x}  A^{\rm (t)}(t)\right)}^2\\
-e\varphi ({\bf r},t)+\hat{v }_{\rm ion}+{v}_{\rm xc}({\bf r},t)\Big]u_{b, {\bf k}}({\bf r},t),
\end{split}
\end{equation} 
where $\varphi ({\bf r},t)$ includes the Hartree potential from the electrons and the local part of the ionic pseudopotentials. 
Here, $\hat{v }_{\rm ion}$ and ${v }_{\rm xc}({\bf r},t)$ are the nonlocal part of the ionic pseudopotentials and exchange-correlation potential, respectively.
The Bloch orbitals $ u_{b,{\bf k}}({\bf r},t)$ are defined in a box containing the unit cell of the 2D thin layer sandwiched by vacuum regions.
The 2D current density $J_{\rm 2D}\left[A^{\rm (t)}\right](t) $ in Eq.~(\ref{GrindEQ__5_}) is derived from the Bloch orbitals as follows:
\begin{equation}
\begin{split}
{\bf J}_{\rm 2D}(t) = -\frac{e}{m} \int dz \int_{\Omega} \frac{dx dy}{\Omega} \sum^{\rm occ}_{b,{\bf k}}
u_{b, {\bf k}}^{\ast}({\bf r},t)  \\
\times \left[ 
-i\hbar \nabla +\hbar{\bf k} + \frac{e}{c} \hat{\bf x}  A^{\rm (t)}(t) + \frac{[{\bf r},\hat{v }_{\rm ion}]}{i\hbar}
\right]
u_{b, {\bf k}}({\bf r},t) ,
\end{split}
\end{equation}
where $\Omega$ is the area of the unit cell and the sum is taken over the occupied orbitals in the ground state.
In the 2D approximation method, coupled  Eq.~(\ref{GrindEQ__5_})  and Eq.~(\ref{GrindEQ__4_}) are simultaneously solved in real time.

TDDFT calculations are performed using SALMON.
The lattice constant of WSe$_{2}$ monolayer is set to $a = b =$ 3.32 {\AA}. The adiabatic local density approximation with Perdew-Zunger functional~\cite{perdew1992accurate} is used for the exchange-correlation.
A slab approximation is used for the $z$ axis with the distance of 20 {\AA} between the atomic monolayers.
The dynamics of the 24 valence electrons are treated explicitly while the effects of the core electrons are considered through norm-conserving pseudopotentials from the OpenMX library~\cite{morrison1993nonlocal}. The spatial grid sizes and k-points are optimized according to the converge results. The determined parameter of the grid size is 0.21 {\AA} while the optimized k-mesh is 15 $\mathrm{\times}$ 15 in the 2D Brillouin zone.

\subsection{Two-band model}

In order to understand the physical mechanism behind the
TDDFT results, we perform model calculations using a minimal band
model~\cite{xiao2012coupled, berkelbach2015bright, kormanyos2015k}. The model Hamiltonian including the second
order coupling for the low energy physics around $K$ or $K'$ point
is described as below:
\begin{equation} 
H^{\tau}[k]
=
\left( \begin{array}{cc}
\frac{\Delta}{2}    &  \tau a\tilde{t} k   
\\
\tau a\tilde{t} k                 &   -\frac{\Delta}{2}
\end{array} \right) 
+
a^2 k^2\left( \begin{array}{cc}
\gamma_1    &  \gamma_3 
\\
\gamma_3   &   \gamma_2
\end{array} \right) 
+
\left( \begin{array}{cc}
0   &  0
\\
0    &  \tau s \lambda 
\end{array} \right) ,
\label{model_hamil} 
\end{equation}
where $\tau=+1\, (-1)$ corresponds to the $K\, (K')$ point, $a$ is the lattice constant,
$\Delta$ is the bandgap, $\tilde{t}$ is the hopping parameter, $\lambda$ is the spin-orbit splitting of the valence band
and  $k$ is  relative to $\tau K$. 
Here we consider only the electron motion along the $x$ axis and omit the $y$ direction ($k_y=0$).
The parameters $\gamma_1$ and $\gamma_2$ represent the breaking of the electron-hole symmetry.
The parameter $\gamma_3$ is responsible for the band asymmetry.
These parameters are determined by fitting the calculated band structure by SALMON.
The first and second terms are the massive Dirac Hamiltonian and its second order correction term, respectively.
The third term is the spin-orbit coupling Hamiltonian and $s=\pm 1$ is the spin index.

By diagonalizing the Hamiltonian, we have the conduction and valence wavefunctions at the ground state:
\begin{equation}
\phi^{\tau}_{{\rm c} k}
=
\left( \begin{array}{cc}
\sqrt{ \frac{\Omega^{\tau}_k+\alpha^{\tau}_k}{2 \Omega^{\tau}_k} } 
\\
s^{\tau}_k \sqrt{ \frac{\Omega^{\tau}_k-\alpha^{\tau}_k}{2 \Omega^{\tau}_k} }        
\end{array} \right) 
,\quad
\phi^{\tau}_{{\rm v}k}
=
\left( \begin{array}{cc}
\sqrt{ \frac{\Omega^{\tau}_k-\alpha^{\tau}_k}{2 \Omega^{\tau}_k} } 
\\
-s^{\tau}_k \sqrt{ \frac{\Omega^{\tau}_k+\alpha^{\tau}_k}{2 \Omega^{\tau}_k} }        
\end{array} \right) ,
\end{equation}
where $\alpha^{\tau}_k = (H^{\tau}_{11}[k]-H^{\tau}_{22}[k])/2$ and $\Omega^{\tau}_{k}=\sqrt{ (\alpha^{\tau}_{k})^2 + (H^{\tau}_{12}[k])^2 }$.
Here $s^{\tau}_k= {\rm sgn}\, H^{\tau}_{12}[k]$ is the sign factor of the off-diagonal element.

The electron dynamics in the presence of the electric field $E(t)$ is described by
\begin{equation}
i\hbar \frac{d}{dt}\psi^{\tau}_k(t) = H^{\tau}\left[k+\frac{e}{\hbar c}A(t)\right] \psi^{\tau}_k(t),
\end{equation}
where $\psi^{\tau}_k(t)=( \psi^{\tau k}_1(t), \psi^{\tau k}_2(t) )^T$ is the time-dependent wavefunction and $A(t)$ is the vector potential (it satisfies  $E(t)=-(1/c)dA(t)/dt$).
The initial value of the wavefunction is taken as follows:
\begin{equation}
\psi^{\tau}_k(t=0) = \phi^{\tau}_{{\rm v} k}.
\end{equation}
The excitation probability from the valence band to the conduction band is written as 
\begin{equation}
P^{\tau}(t) =  \frac{1}{N_k} \sum_{k} \left| \langle \phi^{\tau}_{{\rm c} k} | \psi^{\tau}_k(t) \rangle \right|^2,
\label{Eq:10}
\end{equation} 
where the sum is taken over a certain region of the $k$-point sampling around $k=0$. 
$N_k$ is the number of the sampling points.

To get an intuitive understanding for the transition, we describe below an approximate evaluation of Eq.~(\ref{Eq:10}) using the Landau--Zener theory ignoring the second and third terms in the Hamiltonian of Eq.~(\ref{model_hamil}). We note that the Landau--Zener theory is a semi-classical approximation and can be justified when the applied electric field is sufficiently strong. The valence and conduction eigenenergies are given by
\begin{equation}
\varepsilon_{{\rm c}k}=+\Omega(k), 
\quad 
\varepsilon_{{\rm v}k}=-\Omega(k),
\quad
\Omega(k) \equiv \sqrt{ \frac{\Delta ^2}{4} + a^2 \tilde{t}^2 k^2 }.
\end{equation}
In this paper, we utilize a single-cycle pulse for all the calculations.
In such cases, we can assume that the Landau--Zener transitions occur twice  at the times $t_1$ and $t_2$ ($t_2 > t_1$) for each $k$ point, at which the vector potential crosses $k$ as follows:
\begin{equation}
k+\frac{e}{\hbar c}A(t_{1,2}) = 0.
\end{equation}
The tunnelling probability at $t_1$ and $t_2$ may be written as $P_{\rm LZ}$ and $1-P_{\rm LZ}$, respectively, where 
\begin{equation}
P_{\rm LZ} = \exp \left(-2\pi \frac{\Delta^2}{4\hbar v} \right), \quad 
v=\frac{2e}{\hbar}\left|E(t_1)\right|a \tilde{t}.
\end{equation}
By considering the phase factor due to the adiabatic time evolution, we have
\begin{equation}
P(t)\sim
4 P_{\rm LZ} (1-P_{\rm LZ})
\sin^{2} \left[ \int_{t_{1}}^{t_{2}}dt \, \Omega \left(k+\frac{e}{\hbar c}A(t) \right) \right].
\label{GrindEQ__11_}
\end{equation}
The interference that originates from two transitions at $t_1$ and $t_2$ is known as the Landau-Zener-St\"uckelberg interference.
The cancellation condition by the Landau-Zener-St\"uckelberg
interference is given as, 
\begin{equation}
\int_{t_{1}}^{t_{2}}dt\Omega \left(k+\frac{e}{\hbar c}A(t) \right)=0
\label{GrindEQ__12_}.
\end{equation}

\section{RESULTS AND DISCUSSION}

\begin{figure*}
\centering \includegraphics[scale=0.4]{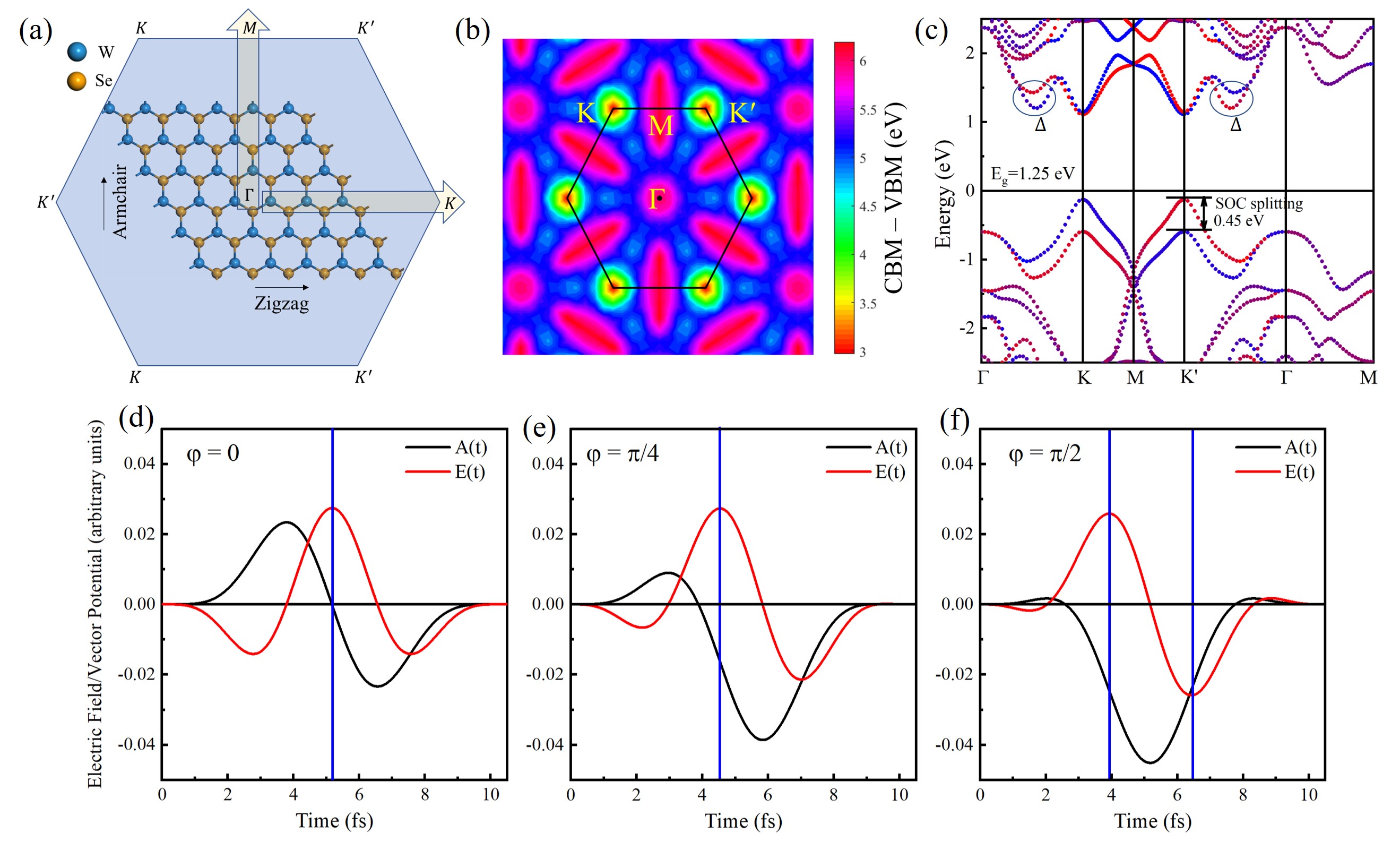} \caption{ (a) WSe$_{2}$ monolayer along with the first BZ. The relevant polarization
directions, armchair ($\mathit{\Gamma}$-$M$ in reciprocal space),
and zigzag ($\mathit{\Gamma}$-$K$) are labeled by arrows. (b) 2D
WSe$_{2\ }$energy map of (CBM-VBM) as a function of the wavevector
$k$ in the hexagonal BZ, and (c) WSe$_{2}$ band structure along high
symmetry directions. The red and blue dots correspond to S$_{z}$
= $\mathrm{\uparrow}$ and S$_{z}$ = $\mathrm{\downarrow}$ respectively.
Electric field and vector potential of single-cycle 10~fs long linearly
polarized laser dependence at various CEP, (d)$\ \varphi=0\mathrm{\ }$
(e)$\ \varphi=\frac{\pi}{4}\mathrm{\ }$and (f)$\ \varphi=\frac{\pi}{2}$.}
\label{fig:1} 
\end{figure*}

 Fig.~\ref{fig:1}(a) shows the monolayer of WSe$_{2}$ along with the
first Brillouin zone (BZ). WSe$_{2}$ is a layered structure where
W atoms are sandwiched between the top and bottom Se layers in a hexagonal
lattice. The six corners of hexagonal BZ contain two inequivalent
high symmetry $K$ and $K^{'}$points at the edges owing to the honeycomb
crystal structure. The real space armchair direction of WSe$_{2}$
belongs to $\mathit{\Gamma}$-$M$ while zigzag
correspond to $\mathit{\Gamma}$-$K$ in reciprocal space. Fig.~\ref{fig:1} (b) shows the dispersion
of the bands (valence band maximum (VBM) - conduction band minimum (CBM)) as a function of the wavevector $k$ in the
whole BZ. The band contour at $K$ ($K^{'}$) points is a triangle
and this so-called trigonal warping indicates the anisotropic carrier
distributions in the WSe$_{2}$ monolayer. The electronic band structure
of the WSe$_{2}$ is shown in Fig.~\ref{fig:1}(c). WSe$_{2}$ has a direct bandgap
of 1.25 eV at $K$ ($K^{'}$) and due to the lack of inversion symmetry,
all bands are split by the intrinsic SOC except at the time-reversal
invariant $\mathit{\Gamma}$ and $M$ point. Thus, owing to time-reversal
symmetry and strong SOC, the top of the valence band of WSe$_{2}$
is spin up (spin down) in the $K$ ($K^{'}$) valley. The energy degenerate
valleys have a large VBM spin splitting of $\mathrm{\sim}$0.45 eV
that agrees well with previous studies ~\cite{zhu2011giant,xiao2012coupled,yuan2013zeeman}.

\begin{figure*}
\centering \includegraphics[scale=0.4]{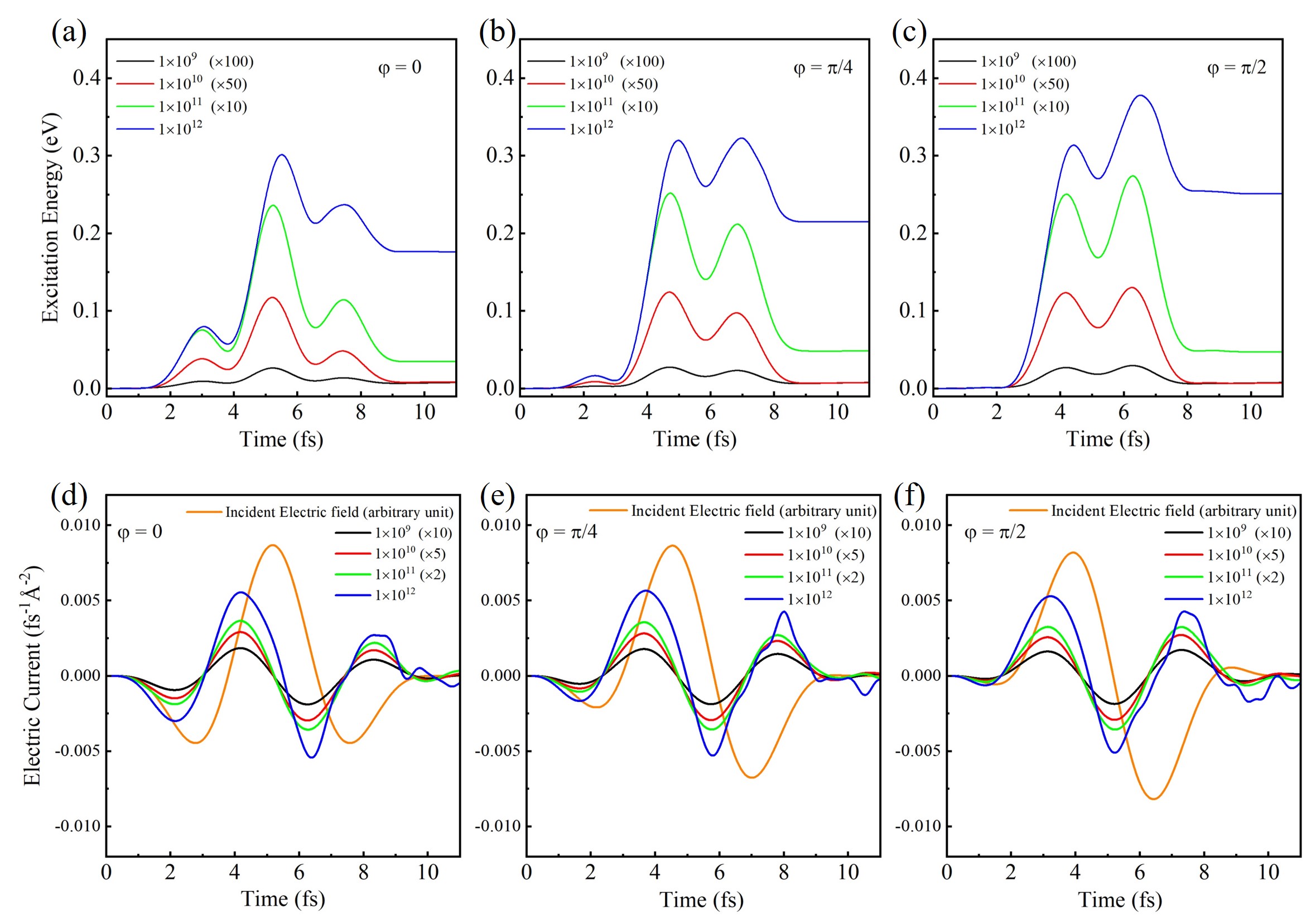} \caption{Temporal development of excitation energies for various intensities
at CEP (a)$\ \varphi=0\mathrm{\ }$ (b)$\ \varphi=\frac{\pi}{4}\mathrm{\ }$and
(c)$\ \varphi=\frac{\pi}{2}$. Applied pulsed electric field and Induced
electric current density as a function of time at CEP (d)$\ \varphi=0\mathrm{\ }$
(e)$\ \varphi=\frac{\pi}{4}\mathrm{\ }$and (f)$\ \varphi=\frac{\pi}{2}$.
For a clear comparison, the results of weak pulses are rescaled up
by multiplying with numerous factors. }
\label{fig:2} 
\end{figure*}

 Long pulses containing several optical cycles resemble a continuous
wave where maxima of electric field concurrence with the zero of the
vector potential. In contrast, ultrashort pulses containing few optical
cycles, the condition (maxima of $E(t)=0$ of $A(t)$) can be controlled
by CEP ($\varphi$). $\varphi$ is the relative phase of the pulse
envelope and the oscillating electric field which plays a significant
role in the pulse waveform for ultrashort laser pulses. Thus, we are
taking advantage of $\varphi$ to manipulate the valley polarization
by using a laser pulse of single optical cycle. To explore the $\varphi$
dependence on valley pseudospin, we apply linearly polarized pulses
parallel to armchair ($\mathit{\Gamma}$-$M$) and zigzag ($\mathit{\Gamma}$-$K$)
directions. We use the vector potential of the following waveform,
\begin{equation}
A^{(i)}\left(t\right)=-\frac{cE_{\rm max}}{\omega}f\left(t\right){\cos}\left\{\omega  \left(t - \frac{T_P}{2} \right)+\varphi\right\}  
\label{GrindEQ__8_}
\end{equation}
where $\omega$ is the average frequency, $E_{\rm max}$ is the maximum
amplitude of the electric field,$\ \varphi$ is CEP, and $T_P$ is the pulse duration. The pulse
envelope function is of $\cos^{4}$ shape for the vector potential
given as 
\begin{equation}
\begin{aligned}f(t)= & \begin{cases}
\cos^{4}\left(\pi\frac{t-T_{P}/2}{T_{P}}\right) & 0\leq t\leq T_{P}\\
0 & \mathrm{otherwise}
\end{cases}\;.\end{aligned}
\label{GrindEQ__9_}
\end{equation}
We use the frequency of 0.4~eV, the pulse length is set to  $T_P=10$~fs and the total computation time is twice the pulse length,
and the time step size is set to 5$\mathrm{\times}$10$^{-4}$~fs.
Fig.~\ref{fig:1}(d-f) shows the electric field and vector potential dependence
on $\varphi$. At $\varphi=0$, the peak of the electric field coincides
with the zero of vector potential. For $\varphi=\frac{\pi}{4}\mathrm{\ }$
vector potential has a nonzero value at the peak of the pulse envelope
while at $\varphi=\frac{\pi}{2}\mathrm{\ }$ the electric field has
positive and negative peaks with the nonzero value of vector potential. 

 We start from the field polarized along the $\mathit{\Gamma}$-$M$
direction. The band contour along with the electronic band structure
(see Fig.~\ref{fig:1}(b, c) displays no asymmetry in $\mathit{\Gamma}$-$M$
direction. Thus, we do not expect valley polarization for the field
polarized along the $\mathit{\Gamma}$-$M$ direction. Note that for
confirmation, valley population at the end of a single-cycle is checked
at different $\varphi$, the valley population is found to be indifferent
at both $K$ and $K^{'}$valley (not shown here). Hence valley polarization
does not exist for the field polarized along $\mathit{\Gamma}$-$M$
because of the lattice symmetry in that direction. On the other hand,
owing to trigonal wrapping, the polarization parallel to $\mathit{\Gamma}$-$K$
experiences different band curvature with respect to $K$ and $K^{'}$point.
Hence, the field polarized along the $\mathit{\Gamma}$-$K$ may experience
strong asymmetries that can lead to the possibility of generating
valley polarization. Therefore, from now on we will focus on the $\mathit{\Gamma}$-$K$
direction.

 Fig.~\ref{fig:2}(a-c) shows the excitation energy for various laser
intensities at $\varphi=0,\ \frac{\pi}{4}\mathrm{,\ and\ }\frac{\pi}{2}$
respectively.
For weak intensity (10$^{9}$ and
10$^{10}$ W/cm$^{2}$), the excitation energy is pronounced during
the irradiation of the pulsed electric field and turns out to be zero
as soon as the pulse ends because the electronic state goes back to
its ground state. On the other hand, the excitation energy at the
intense field ($\mathrm{>}$ 10$^{10\ }$W/cm$^{2}$) is substantially
large and does not vanish even after the pulse ends. Excitation energy
has a more interesting dependence on the $\varphi$ that is independent
of the laser intensity. For $\varphi=0\ $and $\frac{\pi}{4}$ the
electric field has one maxima that are present in the first half cycle
of the pulse thus the excitation energy is dominant in the first half
and reduces in the other half cycle. In contrast, the electric field
at $\varphi=\frac{\pi}{2}\mathrm{\ \ }\ $has two field maxima (positive
and negative) and the excitation energy is even higher in the second
half than the first half cycle. In addition, at the given intensity
the total excitation energy has the order of $\varphi=\frac{\pi}{2}$
$\mathrm{>}$ $\varphi=\frac{\pi}{4}$ $\mathrm{>}$ \ $\varphi=0$.

Before going to the detailed discussion on valley
polarization, the time profile of the incident electric field and
the induced electric current density at multiple \ensuremath{\varphi}
is shown in Fig.~\ref{fig:2}(d-f). The current density depends on the
electric field amplitude as well as on the \ensuremath{\varphi}.
But regardless of the field amplitude and \ensuremath{\varphi}, the
current is not in phase with the incident electric field representing
the typical semiconducting optical response of WSe$_{2}$ monolayer.
The current density at the weak electric field (I = 10$^{9}$ W/cm$^{2}$)
indicates the linear optical response due to a similar time profile
to the pulsed electric field. As the field amplitude increases, the
current starts to depart from the linear response and the distortion
in current density becomes very visible at the intensity I = 10$^{12}$
W/cm$^{2}$, an indication of the strong nonlinear response of electrons.
The behavior of the excitation energy and current density indicates
that valley asymmetry will have a strong dependence on the intensity
and \ensuremath{\varphi}.

\begin{figure*}
\centering \includegraphics[scale=0.3]{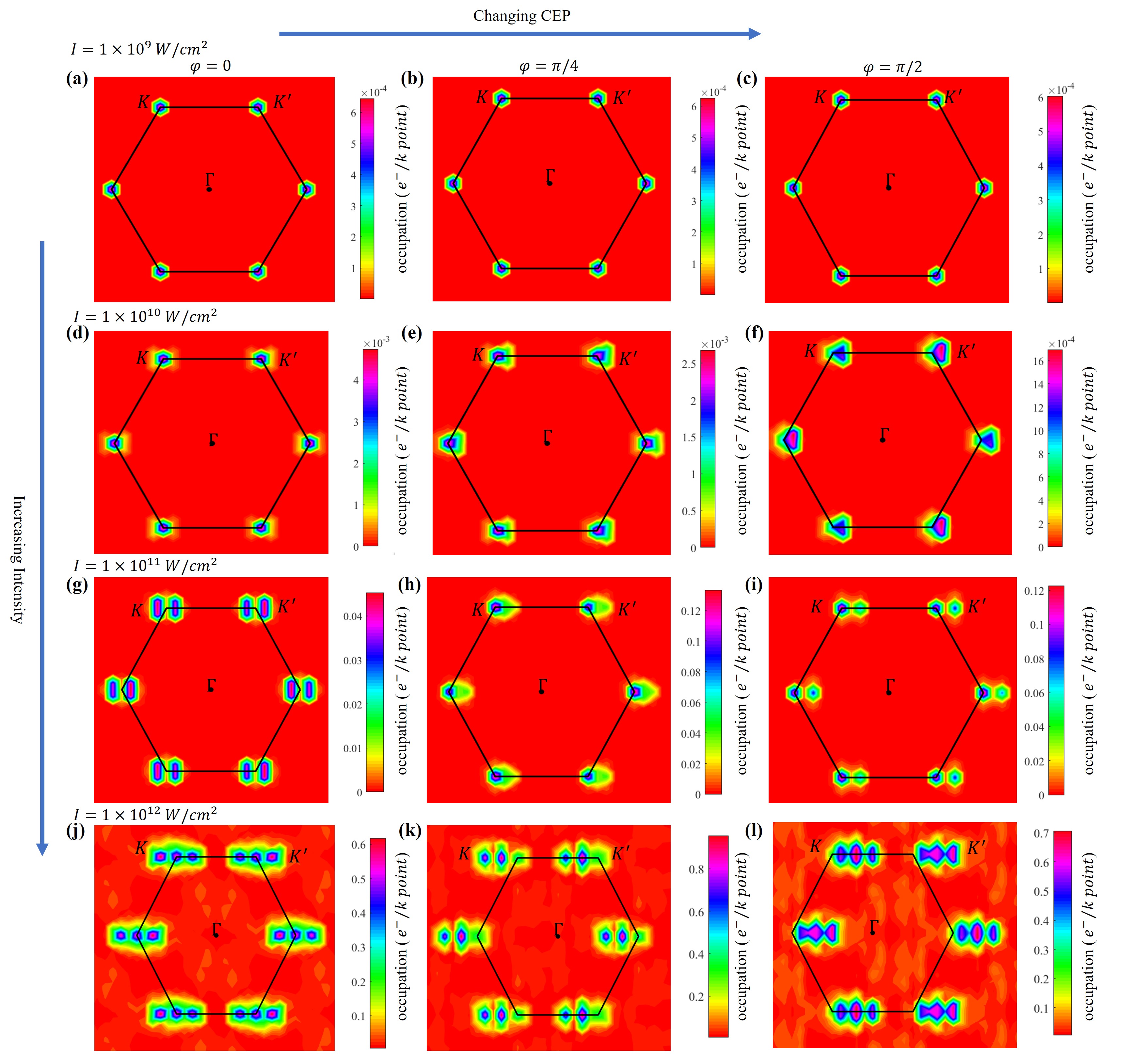} \caption{ Distribution of k-resolved electron populations in the first BZ of
the conduction band at the end of a single optical cycle pulse. Electron
population is summed over the entire conduction band, at various CEP
and intensities. Electron population along $\Gamma$-K direction at
I = 1$\mathrm{\times}$10$^{9}$ W/cm$^{2\ }$field for CEP (a)$\ \varphi=0\mathrm{\ }$
(b)$\ \varphi=\frac{\pi}{4}\mathrm{\ }$and (c)$\ \varphi=\frac{\pi}{2}$.
(d--f) is the same as~(a--c) for I = 1$\mathrm{\times}$10$^{10}$
W/cm$^{2}$, (g--i) for I = 1$\mathrm{\times}$10$^{11}$ W/cm$^{2\ }$while
(j--l) is for I = 1$\mathrm{\times}$10$^{12}$ W/cm$^{2}$.$^{\ }$ }
\label{fig:3} 
\end{figure*}

 We further investigate the distribution of k-resolved electron
populations of the conduction band. Valley population has been shown
in Fig.~\ref{fig:3} at various intensities and $\varphi$ at the end of the
pulse. First, we discuss the effect of intensity on the valley population.
Starting from a very weak intensity of 1$\mathrm{\times}$10$^{9}$
W/cm$^{2}$, we find an equal population at $K$ and $K^{'}$point,
moreover, $\varphi$ dependence is also not realized in the valley
population. Hence no valley asymmetry is present within the limit
of linear optics. As we increase the intensity to 1$\mathrm{\times}$10$^{10}$
W/cm$^{2}$, the difference in the population at $K$ and $K^{'}$point
starts to arise. Further increase in intensity not only increases
the difference in the population at two valleys but also the carrier
density starts to shift around $K$ and $K^{'}$points. 

The intensity
dependence can be understood in a simple manner, as the laser interacts
with the WSe$_{2}$ monolayer, electrons start to tunnel from VBM
to CBM. At weak intensity, the tunneling from VBM to CBM is very weak
and it becomes stronger with the increase in intensity.
The formation of nodes in carrier density distribution
at intense laser fields is observed around $K$($K^{'}$) point. Furthermore,
the difference in valley population also strongly depends on $\varphi$.
As we described above, at $\varphi=0$, vector potential is zero at
the maxima of the electric field which leads the laser field couple
equally to both valleys regardless of the intensity. As the $\varphi$
is varied, the value of the vector potential at the field peaks changes
that control the population difference between two valleys. The results
of the k-resolved population reveal that the valley asymmetry by linearly
polarized pulses is a nonlinear optical phenomenon.

\begin{figure*}
\centering \includegraphics[scale=0.35]{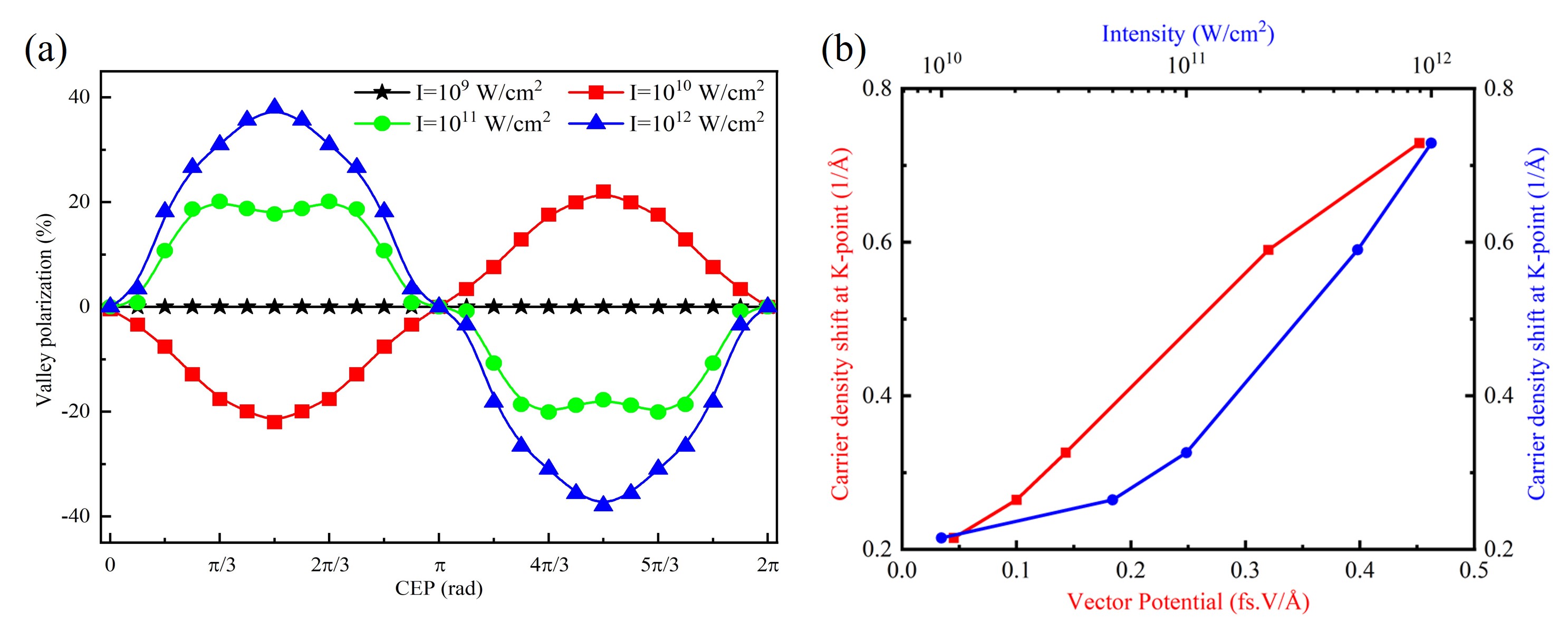} \caption{(a) Valley polarization as a function of CEP at multiple intensities.
(b) Shift in the carrier density as a function of the vector potential
amplitude and intensity. }
\label{fig:4} 
\end{figure*}

 To explore the valley asymmetry as a function of $\varphi$,
we have calculated the valley polarization, shown in \ref{fig:4} (a). Valley
polarization is defined as, 
\begin{equation}
P_{V} \mathrm{=\ 2}\mathrm{\ }\frac{{\rho}_{n,K}-{\rho}_{n,K^{'}}}{{\rho}_{n,K}+{\rho}_{n,K^{'}}}
\label{GrindEQ__10_}
\end{equation}
where~${\rho}_{n,K}({\rho}_{n,K^{'}})$ is obtained by
integrating the electron population in a so-called triangle area whose
size corresponds to the same spin area around $K$($K^{'}$) point.
Note that electron population switches to the opposite valley when
the vector potential changes its sign from negative to positive~for
$\varphi>\pi$. The valley polarization for weakest intensity I =
1$\mathrm{\times}$10$^{9}$ W/cm$^{2}$ is zero for all CEP confirm
the fact that the valley polarization is absent within the linear-optical
limit. By increasing the intensity, we enter in the nonlinear regime
and the substantial valley polarization is observed for I = 1$\mathrm{\times}$10$^{10}$
W/cm$^{2}$. The valley polarization increases gradually with $\varphi$
and reaches its maximum value at $\varphi=\frac{\pi}{2},$ shows a
typical sine wave curve. The valley polarization increases more and
also inverted its sign with the further increase in intensity. The
maximum valley polarization is achieved for the strongest intensity
of 1$\mathrm{\times}$10$^{12}$ W/cm$^{2}$ and the valley polarization
is almost twice as compared to 1$\mathrm{\times}$10$^{10}$ W/cm$^{2}$.
Valley polarization is robust against field strength but
in all cases oscillate as a function of CEP. Although, the valley
polarization is much smaller than the one-photon optical excitation
with circularly polarized pulses nonetheless a sine-like curve as
shown in Fig.~\ref{fig:4}(a) indicates that the valley polarization induced
by linearly polarized pulse can be realized experimentally.

As shown in Fig.~\ref{fig:3}, the carrier density starts to shift around $K$($K^{'}$)
point and the carrier density distribution starts to forms nodes at
strong field intensities. This laser intensity dependence indicates that the position of the carrier density in the BZ can be
controlled with laser intensity. Fig.~\ref{fig:4}(b) shows the
shift in the carrier density as a function of the vector potential
amplitude and intensity at \ensuremath{\varphi} = \ensuremath{\pi}/2
. The vector field amplitude has a linear dependence on the shift
of carrier density. By increasing the intensity, more nodes in the
carrier population start to appear and this may refer to stronger
quantum interferences of wave packets. This will be explained in detail
in the last section.

\begin{figure*}
\centering \includegraphics[scale=0.22]{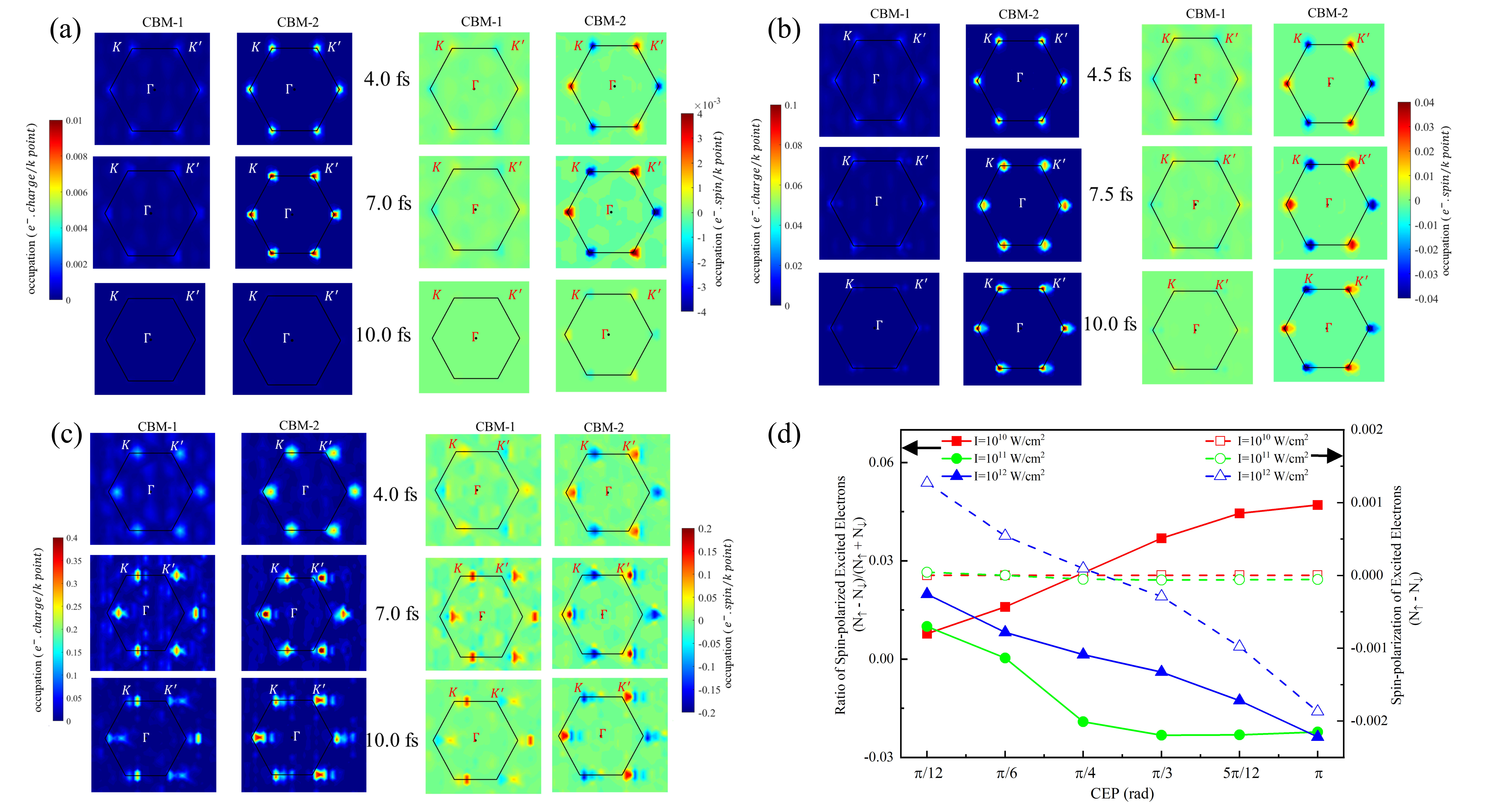} \caption{Charge and spin-decomposed carrier population of concerned conduction
bands named as CBM-1 and CBM-2 at different time steps for (a) I =
1$\mathrm{\times}$10$^{10}$ W/cm$^{2}$, (b) I = 1$\mathrm{\times}$10$^{11}$
W/cm$^{2}$ and (c) I = 1$\mathrm{\times}$10$^{12}$ W/cm$^{2}$.
(d) Spin polarization
of excited charge carriers as a function of CEP.}
\label{fig:5} 
\end{figure*}

 To go through the valley polarization details, we have
drawn the band resolved charge and spin-decomposed carrier population.
Fig.~\ref{fig:5}(a) shows the temporal evolution of charge and spin-resolved
population of intensity 1$\mathrm{\times}$10$^{10}$ W/cm$^{2}$
at $\varphi=\frac{\pi}{2}$. The main concerned bands involved in
this process are CBM-1 and CBM-2 that represent the spin-orbit splitted lower and upper
energy conduction band respectively. Three-time steps are chosen
as, around the first and second maxima of the electric field and at
the end of the pulse. CBM-2 has the same spin as VBM thus at 4.0~fs
the electrons are excited to CBM-2 at $K$($K^{'}$). One can see
that at 7.0~fs which is the second half of the pulse, more electrons
are excited and we note the asymmetry in the population at that point.
The excited electron population becomes small when the laser field ends at 10~fs because most of the electrons go back to their ground state due to very weak intensity. Spin is also confined
at $K$ and $K^{'}$points. The charge and spin-resolved population
of intensity 1$\mathrm{\times}$10$^{11}$ W/cm$^{2}$ is shown in

Fig.~\ref{fig:5}(b) at $\varphi=\frac{\pi}{4}$ (where we find the maximum valley
asymmetry). Charge and spin dynamics are the same as found at 1$\mathrm{\times}$10$^{10}$
W/cm$^{2}$ intensity nevertheless the excited electrons reside in
the CBM-2 even the pulse ends. At the most intense case of the 1$\mathrm{\times}$10$^{12}$
W/cm$^{2}$ field, we find highly nonlinear interaction and electrons
spread more widely throughout the Brillouin zone as shown in Fig.~\ref{fig:5}(c). 
The delta$\ \left(\Delta\right)$ points at CBM close to $K$($K^{'}$)
with opposite spin (see  Fig.~\ref{fig:1}(c)) plays an important role at this
strong intensity. $\Delta$-point acts as an intermediate point which
facilitates the intervalley transfer of excited electrons to lower
and upper conduction bands. Thus, unlike the other cases, the charge
along with the spin is not limited to CBM-2 and multiple conduction
bands start to contribute in valley polarization.

The spin polarization of excited charge carriers is shown in Fig. \ref{fig:5}(d).
Overall the spin polarization (N$_{\uparrow}$
- N$_{\downarrow}$) is negligible and independent of intensity and \ensuremath{\varphi},
except a minute spin starts to appear at I = 1$\mathrm{\times}$10$^{12}$
W/cm$^{2}$.
Degree of spin polarization (N$_{\uparrow}$- N$_{\downarrow}$)/(N$_{\uparrow}$+ N$_{\downarrow}$) follow the same behavior as valley polarization
which shows that spin polarization is also an observable along with valley polarization.

\begin{figure*}
\centering \includegraphics[scale=0.22]{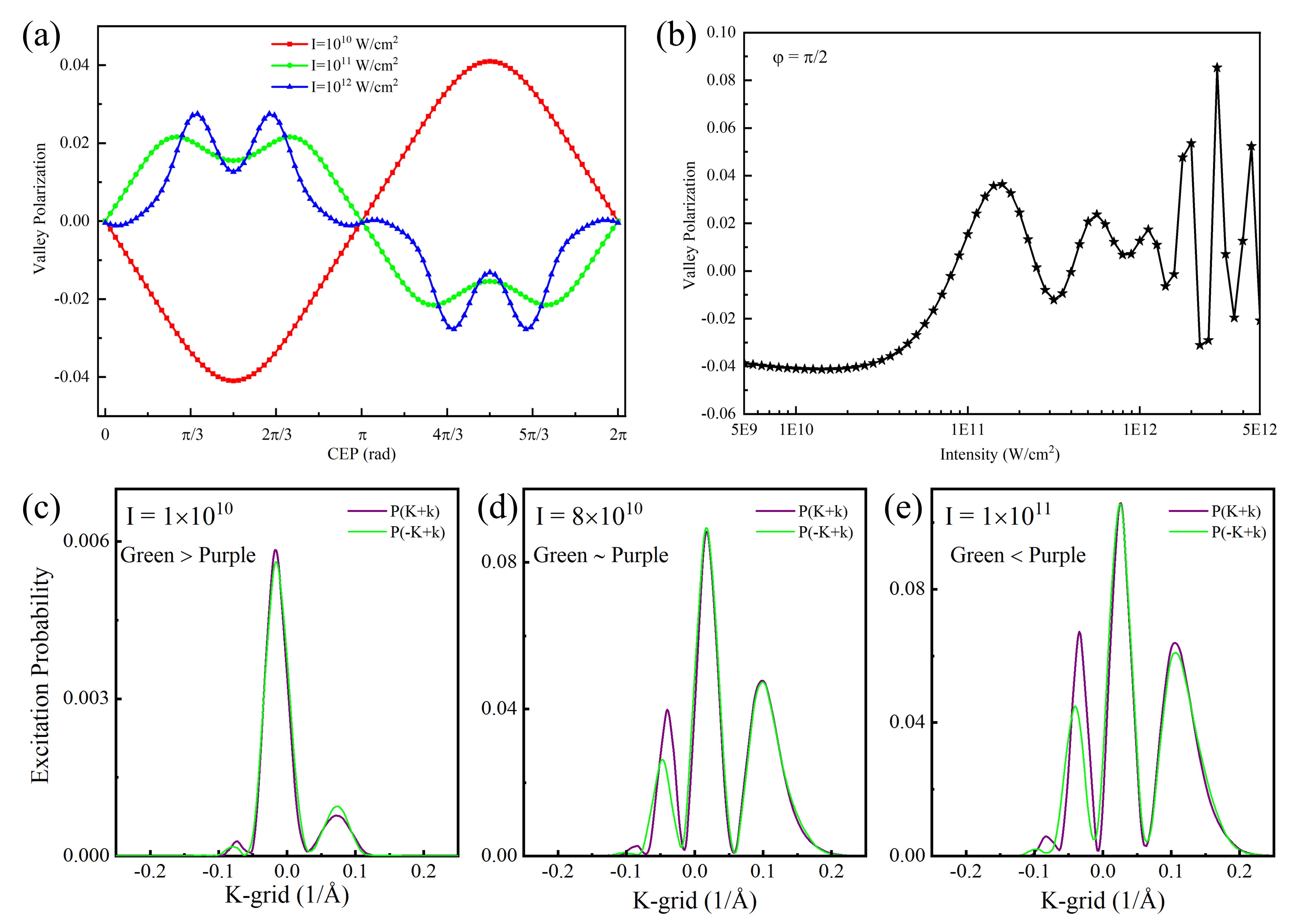} \caption{Massive Dirac Hamiltonian two-band model results (a) Valley polarization
as a function of CEP at multiple intensities. (b) Valley polarization
as a function of intensity at $\varphi=\pi/2$. Excitation probabilities
at (c) 1$\mathrm{\times}$10$^{10}$ W/cm$^{2}$, (d) 8$\mathrm{\times}$10$^{10}$
W/cm$^{2}$ and (e) 1$\mathrm{\times}$10$^{11}$ W/cm$^{2}$.}
\label{fig:6} 
\end{figure*}

The valley polarization results calculated by the two-band model as
a function of \ensuremath{\varphi} are shown in  
Fig.~\ref{fig:6}(a). Two band
model qualitatively reproduces the overall trends of valley polarization
with TDDFT results. The origin of the carrier density nodes at higher
intensities and the phase change of the valley polarization with respect
to intensity can be explained by the two-band model. 
Based on the massive Dirac model, the phase appearing in the Landau--Zener formula is referred to St\"uckelberg phase as described in Eq.~(\ref{GrindEQ__11_}) and Eq.~(\ref{GrindEQ__12_}).
The recurring Landau--Zener transitions drive by an oscillatory external
field produces an excitation density matrix with an opposite sign
which causes the interference of the wave packets. Thus the carrier
density nodes appearing at high intensity may refer to the St\"uckelberg
interference where the excitation probability becomes zero at a certain
$k$. 

Fig.~\ref{fig:6}(b) shows the intensity dependence of the valley polarization
at $\varphi=\frac{\pi}{2}$. Valley polarization has a very complex
behavior regarding the phase change. Thus to find the origin of the
phase change, we have shown the excitation probability in Fig.~\ref{fig:6}(c-e).
The excitation probability of P(K+k) \textgreater{} P(-K+k) at 1$\mathrm{\times}$10$^{10}$
W/cm$^{2}$, as we increase intensity, P(-K+k) starts to increase,
and the excitation probability roughly the same at 8$\mathrm{\times}$10$^{10}$
W/cm$^{2}$. Further increase in intensity results P(K+k) \textless{}
P(-K+k) and this brings the phase change of the valley polarization.
Overall, the St\"uckelberg interference first takes place at the positive
k-region, and asymmetry is observed at the peak caused by the interference.
The peak at the negative $k$-region follows at higher intensities and
causes stronger interference of wave packets. 

\section{CONCLUSION}

 In conclusion, we investigated the single cycle pulse control
of valley pseudospin in the WSe$_{2}$ monolayer. The intensity and
CEP dependence of the pulsed electric field is varied to investigate
the mechanism of valley polarization. Linearly polarized pulse along
armchair and zigzag directions are applied. 

Valley polarization remains
zero within the linear optical limit for both polarization directions.
In the nonlinear regime, no valley asymmetry and its CEP dependency
is realized for the field polarized along the armchair direction while
the polarization parallels to the zigzag direction experience strong
asymmetries. 
The valley polarization is small at weak intensities
but it increases with the increase in intensity and substantial valley
polarization is achieved. The valley polarization is robust
against field strength but it strongly depends on the CEP.

We showed
that in the strong-field regime the electron dynamics display quantum
interference that gives rise to distinct node formation. More importantly,
the position of the carrier density is strongly dependent on laser
intensity which indicates the possibility to control the electron
momentum in BZ. Two band model indicates that the
carrier density nodes appearing at high intensity may refer to the
Stuckelberg interference. 

Our results provide the opportunity to
manipulate the valley pseudospin and optical field control of electron
dynamics faster than electron-electron scattering and electron-phonon
scattering.  

\section{AUTHOR INFORMATION}
\noindent \textbf{Corresponding Author}

\noindent *E-mail: otobe.tomohito@qst.go.jp  

\noindent \textbf{Author Contributions}


\noindent \textbf{Notes}

\noindent The authors declare no competing financial interest.

\noindent 
\section{ACKNOWLEDGMENT}
This research is supported by JST-CREST under Grant No. JP-MJCR16N5. This research is also partially supported by JSPS KAKENHI Grant No. 20H02649, and MEXT Quantum Leap Flagship Program (MEXT Q-LEAP) under Grant No. JPMXS0118068681. The numerical calculations are carried out using the computer facilities of the Fugaku through the HPCI System Research Project (Project ID: hp210137), SGI8600 at Japan Atomic Energy Agency (JAEA), and Multidisciplinary Cooperative Research Program in CCS, University of Tsukuba.

\bibliographystyle{naturemag}
\bibliography{bibliography}

\begin{thebibliography}{10}
\expandafter\ifx\csname url\endcsname\relax
  \def\url#1{\texttt{#1}}\fi
\expandafter\ifx\csname urlprefix\endcsname\relax\def\urlprefix{URL }\fi
\providecommand{\bibinfo}[2]{#2}
\providecommand{\eprint}[2][]{\url{#2}}

\bibitem{osada2012two}
\bibinfo{author}{Osada, M.} \& \bibinfo{author}{Sasaki, T.}
\newblock \bibinfo{title}{Two-dimensional dielectric nanosheets: novel
  nanoelectronics from nanocrystal building blocks}.
\newblock \emph{\bibinfo{journal}{Advanced Materials}}
  \textbf{\bibinfo{volume}{24}}, \bibinfo{pages}{210--228}
  (\bibinfo{year}{2012}).

\bibitem{chhowalla2013chemistry}
\bibinfo{author}{Chhowalla, M.} \emph{et~al.}
\newblock \bibinfo{title}{The chemistry of two-dimensional layered transition
  metal dichalcogenide nanosheets}.
\newblock \emph{\bibinfo{journal}{Nature chemistry}}
  \textbf{\bibinfo{volume}{5}}, \bibinfo{pages}{263--275}
  (\bibinfo{year}{2013}).

\bibitem{novoselov2004electric}
\bibinfo{author}{Novoselov, K.~S.} \emph{et~al.}
\newblock \bibinfo{title}{Electric field effect in atomically thin carbon
  films}.
\newblock \emph{\bibinfo{journal}{science}} \textbf{\bibinfo{volume}{306}},
  \bibinfo{pages}{666--669} (\bibinfo{year}{2004}).

\bibitem{kang20202d}
\bibinfo{author}{Kang, S.} \emph{et~al.}
\newblock \bibinfo{title}{2d semiconducting materials for electronic and
  optoelectronic applications: potential and challenge}.
\newblock \emph{\bibinfo{journal}{2D Materials}} \textbf{\bibinfo{volume}{7}},
  \bibinfo{pages}{022003} (\bibinfo{year}{2020}).

\bibitem{novoselov2012roadmap}
\bibinfo{author}{Novoselov, K.~S.} \emph{et~al.}
\newblock \bibinfo{title}{A roadmap for graphene}.
\newblock \emph{\bibinfo{journal}{nature}} \textbf{\bibinfo{volume}{490}},
  \bibinfo{pages}{192--200} (\bibinfo{year}{2012}).

\bibitem{tao2015silicene}
\bibinfo{author}{Tao, L.} \emph{et~al.}
\newblock \bibinfo{title}{Silicene field-effect transistors operating at room
  temperature}.
\newblock \emph{\bibinfo{journal}{Nature nanotechnology}}
  \textbf{\bibinfo{volume}{10}}, \bibinfo{pages}{227--231}
  (\bibinfo{year}{2015}).

\bibitem{wang2012electronics}
\bibinfo{author}{Wang, Q.~H.}, \bibinfo{author}{Kalantar-Zadeh, K.},
  \bibinfo{author}{Kis, A.}, \bibinfo{author}{Coleman, J.~N.} \&
  \bibinfo{author}{Strano, M.~S.}
\newblock \bibinfo{title}{Electronics and optoelectronics of two-dimensional
  transition metal dichalcogenides}.
\newblock \emph{\bibinfo{journal}{Nature nanotechnology}}
  \textbf{\bibinfo{volume}{7}}, \bibinfo{pages}{699--712}
  (\bibinfo{year}{2012}).

\bibitem{xia2014rediscovering}
\bibinfo{author}{Xia, F.}, \bibinfo{author}{Wang, H.} \& \bibinfo{author}{Jia,
  Y.}
\newblock \bibinfo{title}{Rediscovering black phosphorus as an anisotropic
  layered material for optoelectronics and electronics}.
\newblock \emph{\bibinfo{journal}{Nature communications}}
  \textbf{\bibinfo{volume}{5}}, \bibinfo{pages}{1--6} (\bibinfo{year}{2014}).

\bibitem{tan20202d}
\bibinfo{author}{Tan, T.}, \bibinfo{author}{Jiang, X.}, \bibinfo{author}{Wang,
  C.}, \bibinfo{author}{Yao, B.} \& \bibinfo{author}{Zhang, H.}
\newblock \bibinfo{title}{2d material optoelectronics for information
  functional device applications: status and challenges}.
\newblock \emph{\bibinfo{journal}{Advanced Science}}
  \textbf{\bibinfo{volume}{7}}, \bibinfo{pages}{2000058}
  (\bibinfo{year}{2020}).

\bibitem{wen2019nonlinear}
\bibinfo{author}{Wen, X.}, \bibinfo{author}{Gong, Z.} \& \bibinfo{author}{Li,
  D.}
\newblock \bibinfo{title}{Nonlinear optics of two-dimensional transition metal
  dichalcogenides}.
\newblock \emph{\bibinfo{journal}{InfoMat}} \textbf{\bibinfo{volume}{1}},
  \bibinfo{pages}{317--337} (\bibinfo{year}{2019}).

\bibitem{zhou20182d}
\bibinfo{author}{Zhou, X.} \emph{et~al.}
\newblock \bibinfo{title}{2d layered material-based van der waals
  heterostructures for optoelectronics}.
\newblock \emph{\bibinfo{journal}{Advanced Functional Materials}}
  \textbf{\bibinfo{volume}{28}}, \bibinfo{pages}{1706587}
  (\bibinfo{year}{2018}).

\bibitem{schaibley2016valleytronics}
\bibinfo{author}{Schaibley, J.~R.} \emph{et~al.}
\newblock \bibinfo{title}{Valleytronics in 2d materials}.
\newblock \emph{\bibinfo{journal}{Nature Reviews Materials}}
  \textbf{\bibinfo{volume}{1}}, \bibinfo{pages}{1--15} (\bibinfo{year}{2016}).

\bibitem{yao2008valley}
\bibinfo{author}{Yao, W.}, \bibinfo{author}{Xiao, D.} \& \bibinfo{author}{Niu,
  Q.}
\newblock \bibinfo{title}{Valley-dependent optoelectronics from inversion
  symmetry breaking}.
\newblock \emph{\bibinfo{journal}{Physical Review B}}
  \textbf{\bibinfo{volume}{77}}, \bibinfo{pages}{235406}
  (\bibinfo{year}{2008}).

\bibitem{zhu2011giant}
\bibinfo{author}{Zhu, Z.~Y.}, \bibinfo{author}{Cheng, Y.~C.} \&
  \bibinfo{author}{Schwingenschl{\"o}gl, U.}
\newblock \bibinfo{title}{Giant spin-orbit-induced spin splitting in
  two-dimensional transition-metal dichalcogenide semiconductors}.
\newblock \emph{\bibinfo{journal}{Physical Review B}}
  \textbf{\bibinfo{volume}{84}}, \bibinfo{pages}{153402}
  (\bibinfo{year}{2011}).

\bibitem{xiao2012coupled}
\bibinfo{author}{Xiao, D.}, \bibinfo{author}{Liu, G.-B.},
  \bibinfo{author}{Feng, W.}, \bibinfo{author}{Xu, X.} \& \bibinfo{author}{Yao,
  W.}
\newblock \bibinfo{title}{Coupled spin and valley physics in monolayers of mos
  2 and other group-vi dichalcogenides}.
\newblock \emph{\bibinfo{journal}{Physical review letters}}
  \textbf{\bibinfo{volume}{108}}, \bibinfo{pages}{196802}
  (\bibinfo{year}{2012}).

\bibitem{yuan2013zeeman}
\bibinfo{author}{Yuan, H.} \emph{et~al.}
\newblock \bibinfo{title}{Zeeman-type spin splitting controlled by an electric
  field}.
\newblock \emph{\bibinfo{journal}{Nature Physics}}
  \textbf{\bibinfo{volume}{9}}, \bibinfo{pages}{563--569}
  (\bibinfo{year}{2013}).

\bibitem{xu2014spin}
\bibinfo{author}{Xu, X.}, \bibinfo{author}{Yao, W.}, \bibinfo{author}{Xiao, D.}
  \& \bibinfo{author}{Heinz, T.~F.}
\newblock \bibinfo{title}{Spin and pseudospins in layered transition metal
  dichalcogenides}.
\newblock \emph{\bibinfo{journal}{Nature Physics}}
  \textbf{\bibinfo{volume}{10}}, \bibinfo{pages}{343--350}
  (\bibinfo{year}{2014}).

\bibitem{zeng2012valley}
\bibinfo{author}{Zeng, H.}, \bibinfo{author}{Dai, J.}, \bibinfo{author}{Yao,
  W.}, \bibinfo{author}{Xiao, D.} \& \bibinfo{author}{Cui, X.}
\newblock \bibinfo{title}{Valley polarization in mos 2 monolayers by optical
  pumping}.
\newblock \emph{\bibinfo{journal}{Nature nanotechnology}}
  \textbf{\bibinfo{volume}{7}}, \bibinfo{pages}{490--493}
  (\bibinfo{year}{2012}).

\bibitem{mak2012control}
\bibinfo{author}{Mak, K.~F.}, \bibinfo{author}{He, K.}, \bibinfo{author}{Shan,
  J.} \& \bibinfo{author}{Heinz, T.~F.}
\newblock \bibinfo{title}{Control of valley polarization in monolayer mos 2 by
  optical helicity}.
\newblock \emph{\bibinfo{journal}{Nature nanotechnology}}
  \textbf{\bibinfo{volume}{7}}, \bibinfo{pages}{494--498}
  (\bibinfo{year}{2012}).

\bibitem{yuan2014generation}
\bibinfo{author}{Yuan, H.} \emph{et~al.}
\newblock \bibinfo{title}{Generation and electric control of
  spin--valley-coupled circular photogalvanic current in wse 2}.
\newblock \emph{\bibinfo{journal}{Nature nanotechnology}}
  \textbf{\bibinfo{volume}{9}}, \bibinfo{pages}{851--857}
  (\bibinfo{year}{2014}).

\bibitem{sanchez2016valley}
\bibinfo{author}{Sanchez, O.~L.}, \bibinfo{author}{Ovchinnikov, D.},
  \bibinfo{author}{Misra, S.}, \bibinfo{author}{Allain, A.} \&
  \bibinfo{author}{Kis, A.}
\newblock \bibinfo{title}{Valley polarization by spin injection in a
  light-emitting van der waals heterojunction}.
\newblock \emph{\bibinfo{journal}{Nano letters}} \textbf{\bibinfo{volume}{16}},
  \bibinfo{pages}{5792--5797} (\bibinfo{year}{2016}).

\bibitem{ye2016electrical}
\bibinfo{author}{Ye, Y.} \emph{et~al.}
\newblock \bibinfo{title}{Electrical generation and control of the valley
  carriers in a monolayer transition metal dichalcogenide}.
\newblock \emph{\bibinfo{journal}{Nature nanotechnology}}
  \textbf{\bibinfo{volume}{11}}, \bibinfo{pages}{598--602}
  (\bibinfo{year}{2016}).

\bibitem{norden2019giant}
\bibinfo{author}{Norden, T.} \emph{et~al.}
\newblock \bibinfo{title}{Giant valley splitting in monolayer ws 2 by magnetic
  proximity effect}.
\newblock \emph{\bibinfo{journal}{Nature communications}}
  \textbf{\bibinfo{volume}{10}}, \bibinfo{pages}{1--10} (\bibinfo{year}{2019}).

\bibitem{vitale2018valleytronics}
\bibinfo{author}{Vitale, S.~A.} \emph{et~al.}
\newblock \bibinfo{title}{Valleytronics: opportunities, challenges, and paths
  forward}.
\newblock \emph{\bibinfo{journal}{Small}} \textbf{\bibinfo{volume}{14}},
  \bibinfo{pages}{1801483} (\bibinfo{year}{2018}).

\bibitem{langer2018lightwave}
\bibinfo{author}{Langer, F.} \emph{et~al.}
\newblock \bibinfo{title}{Lightwave valleytronics in a monolayer of tungsten
  diselenide}.
\newblock \emph{\bibinfo{journal}{Nature}} \textbf{\bibinfo{volume}{557}},
  \bibinfo{pages}{76--80} (\bibinfo{year}{2018}).

\bibitem{jimenez2020lightwave}
\bibinfo{author}{Jim{\'e}nez-Gal{\'a}n, {\'A}.}, \bibinfo{author}{Silva, R.},
  \bibinfo{author}{Smirnova, O.} \& \bibinfo{author}{Ivanov, M.}
\newblock \bibinfo{title}{Lightwave control of topological properties in 2d
  materials for sub-cycle and non-resonant valley manipulation}.
\newblock \emph{\bibinfo{journal}{Nature Photonics}}
  \textbf{\bibinfo{volume}{14}}, \bibinfo{pages}{728--732}
  (\bibinfo{year}{2020}).

\bibitem{jimenez2021sub}
\bibinfo{author}{Jim{\'e}nez-Gal{\'a}n, {\'A}.}, \bibinfo{author}{Silva,
  R.~E.}, \bibinfo{author}{Smirnova, O.} \& \bibinfo{author}{Ivanov, M.}
\newblock \bibinfo{title}{Sub-cycle valleytronics: control of valley
  polarization using few-cycle linearly polarized pulses}.
\newblock \emph{\bibinfo{journal}{Optica}} \textbf{\bibinfo{volume}{8}},
  \bibinfo{pages}{277--280} (\bibinfo{year}{2021}).

\bibitem{otobe2008first}
\bibinfo{author}{Otobe, T.} \emph{et~al.}
\newblock \bibinfo{title}{First-principles electron dynamics simulation for
  optical breakdown of dielectrics under an intense laser field}.
\newblock \emph{\bibinfo{journal}{Physical Review B}}
  \textbf{\bibinfo{volume}{77}}, \bibinfo{pages}{165104}
  (\bibinfo{year}{2008}).

\bibitem{noda2019salmon}
\bibinfo{author}{Noda, M.} \emph{et~al.}
\newblock \bibinfo{title}{Salmon: Scalable ab-initio light--matter simulator
  for optics and nanoscience}.
\newblock \emph{\bibinfo{journal}{Computer Physics Communications}}
  \textbf{\bibinfo{volume}{235}}, \bibinfo{pages}{356--365}
  (\bibinfo{year}{2019}).

\bibitem{Salmon:Online}
\bibinfo{title}{Salmon official}.
\newblock \bibinfo{howpublished}{\url{http://salmon-tddft.jp}}.

\bibitem{von1972local}
\bibinfo{author}{Von~Barth, U.} \& \bibinfo{author}{Hedin, L.}
\newblock \bibinfo{title}{A local exchange-correlation potential for the spin
  polarized case. i}.
\newblock \emph{\bibinfo{journal}{Journal of Physics C: Solid State Physics}}
  \textbf{\bibinfo{volume}{5}}, \bibinfo{pages}{1629} (\bibinfo{year}{1972}).

\bibitem{oda1998fully}
\bibinfo{author}{Oda, T.}, \bibinfo{author}{Pasquarello, A.} \&
  \bibinfo{author}{Car, R.}
\newblock \bibinfo{title}{Fully unconstrained approach to noncollinear
  magnetism: application to small fe clusters}.
\newblock \emph{\bibinfo{journal}{Physical review letters}}
  \textbf{\bibinfo{volume}{80}}, \bibinfo{pages}{3622} (\bibinfo{year}{1998}).

\bibitem{theurich2001self}
\bibinfo{author}{Theurich, G.} \& \bibinfo{author}{Hill, N.~A.}
\newblock \bibinfo{title}{Self-consistent treatment of spin-orbit coupling in
  solids using relativistic fully separable ab initio pseudopotentials}.
\newblock \emph{\bibinfo{journal}{Physical Review B}}
  \textbf{\bibinfo{volume}{64}}, \bibinfo{pages}{073106}
  (\bibinfo{year}{2001}).

\bibitem{yamada2018time}
\bibinfo{author}{Yamada, S.}, \bibinfo{author}{Noda, M.},
  \bibinfo{author}{Nobusada, K.} \& \bibinfo{author}{Yabana, K.}
\newblock \bibinfo{title}{Time-dependent density functional theory for
  interaction of ultrashort light pulse with thin materials}.
\newblock \emph{\bibinfo{journal}{Physical Review B}}
  \textbf{\bibinfo{volume}{98}}, \bibinfo{pages}{245147}
  (\bibinfo{year}{2018}).

\bibitem{yamada2021determining}
\bibinfo{author}{Yamada, S.} \& \bibinfo{author}{Yabana, K.}
\newblock \bibinfo{title}{Determining the optimum thickness for high harmonic
  generation from nanoscale thin films: An ab initio computational study}.
\newblock \emph{\bibinfo{journal}{Physical Review B}}
  \textbf{\bibinfo{volume}{103}}, \bibinfo{pages}{155426}
  (\bibinfo{year}{2021}).

\bibitem{bertsch2000real}
\bibinfo{author}{Bertsch, G.~F.}, \bibinfo{author}{Iwata, J.-I.},
  \bibinfo{author}{Rubio, A.} \& \bibinfo{author}{Yabana, K.}
\newblock \bibinfo{title}{Real-space, real-time method for the dielectric
  function}.
\newblock \emph{\bibinfo{journal}{Physical Review B}}
  \textbf{\bibinfo{volume}{62}}, \bibinfo{pages}{7998} (\bibinfo{year}{2000}).

\bibitem{perdew1992accurate}
\bibinfo{author}{Perdew, J.~P.} \& \bibinfo{author}{Wang, Y.}
\newblock \bibinfo{title}{Accurate and simple analytic representation of the
  electron-gas correlation energy}.
\newblock \emph{\bibinfo{journal}{Physical review B}}
  \textbf{\bibinfo{volume}{45}}, \bibinfo{pages}{13244} (\bibinfo{year}{1992}).

\bibitem{morrison1993nonlocal}
\bibinfo{author}{Morrison, I.}, \bibinfo{author}{Bylander, D.} \&
  \bibinfo{author}{Kleinman, L.}
\newblock \bibinfo{title}{Nonlocal hermitian norm-conserving vanderbilt
  pseudopotential}.
\newblock \emph{\bibinfo{journal}{Physical Review B}}
  \textbf{\bibinfo{volume}{47}}, \bibinfo{pages}{6728} (\bibinfo{year}{1993}).

\bibitem{berkelbach2015bright}
\bibinfo{author}{Berkelbach, T.~C.}, \bibinfo{author}{Hybertsen, M.~S.} \&
  \bibinfo{author}{Reichman, D.~R.}
\newblock \bibinfo{title}{Bright and dark singlet excitons via linear and
  two-photon spectroscopy in monolayer transition-metal dichalcogenides}.
\newblock \emph{\bibinfo{journal}{Physical Review B}}
  \textbf{\bibinfo{volume}{92}}, \bibinfo{pages}{085413}
  (\bibinfo{year}{2015}).

\bibitem{kormanyos2015k}
\bibinfo{author}{Korm{\'a}nyos, A.} \emph{et~al.}
\newblock \bibinfo{title}{k{\textperiodcentered} p theory for two-dimensional
  transition metal dichalcogenide semiconductors}.
\newblock \emph{\bibinfo{journal}{2D Materials}} \textbf{\bibinfo{volume}{2}},
  \bibinfo{pages}{022001} (\bibinfo{year}{2015}).

\end{thebibliography}
\noindent \eject 
\end{document}